\documentstyle[aps,prl,epsfig]{revtex}

\begin{document}

\twocolumn[\hsize\textwidth\columnwidth\hsize\csname @twocolumnfalse\endcsname

\title{Do Chaotic Trajectories Care About Self-Similarity?}

\author{M.~Weiss$^{1,2}$, L.~Hufnagel$^1$, and R.~Ketzmerick$^1$ \\
$^1$ Max-Planck-Institut f\"ur Str\"omungsforschung
and Institut f\"ur Nichtlineare Dynamik der Universit\"at G\"ottingen,\\
Bunsenstr.~10, 37073 G\"ottingen, Germany\\
$^2$ EMBL, Meyerhofstr.~1, 69117 Heidelberg, Germany
}
\maketitle

\begin{abstract}
We investigate the relation between the chaotic dynamics and the hierarchical phase-space
structure of generic Hamiltonian systems.
We demonstrate that even in ideal situations when the phase space is dominated by an exactly
self-similar structure, the long-time dynamics is {\it not} dominated by this structure.
This has consequences for the power-law decay of correlations and Poincar\'e recurrences.
\end{abstract}

\pacs{PACS number: 05.45.Mt}
]

\begin{figure}
\begin{center}
    \epsfxsize=8.4cm
    \leavevmode
    \epsffile{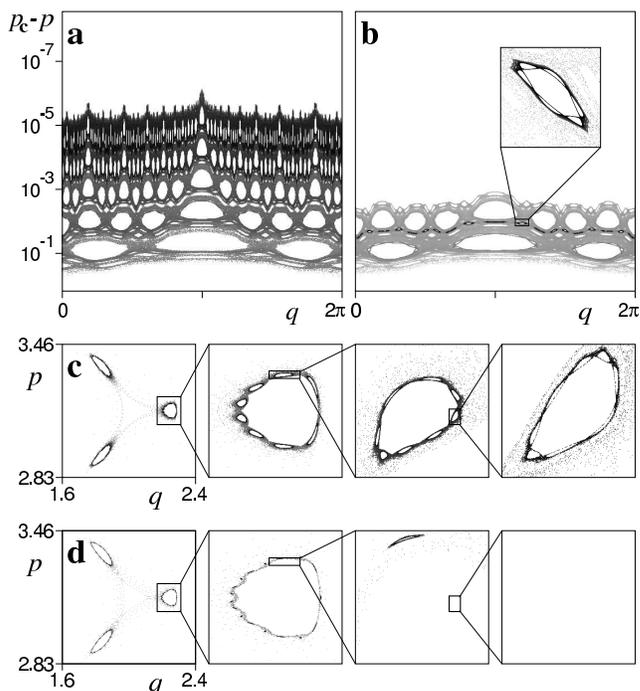}
\caption{
Phase-space density of individual trajectories of length $\approx5\cdot10^7$ 
for the symmetrized standard map.
(a,b) $K=K_c$: By stretching the phase space in $p$ direction according to the distance $p_c(q)-p$
to the critical golden torus $p_c(q)$ the self-similar phase-space structure in its vicinity 
is visualized. The density is determined on a $500\!\times\!500$ grid.
(c,d) $K=K^*$: Successive magnifications of the island-around-island sequence $3-8-8-8\ldots$.
The density is determined on a $250\!\times\!250$ grid.
In (a) and (c) the trajectory follows the self-similar phase-space structure,
while (b) and (d) show representatives of the counter examples.
Grey shadings are on a logarithmic scale.
}
\label{fig:density}
\end{center}
\end{figure}

\begin{figure}
\begin{center}
    \epsfxsize=8.4cm
    \leavevmode
    \epsffile{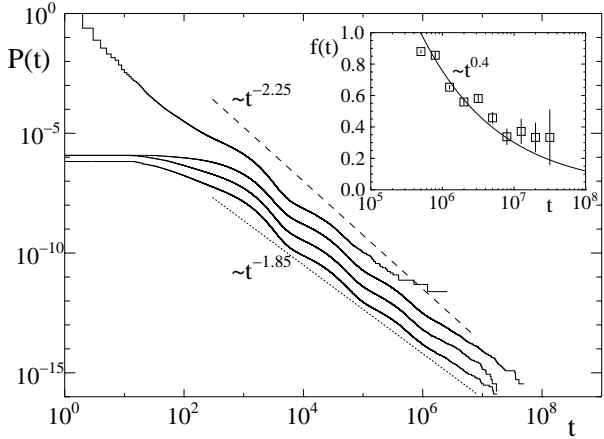}
\caption[fig:zas]{
$P(t)$ for $K=K^*$ for various initial
conditions~\cite{remark1} (vertically shifted for better comparison).
All four curves (solid) show the same power-law behavior including the fluctuations
and deviate from the prediction $\gamma=2.25$ from Ref.~\cite{zaslavsky} (dashed).
The inset shows the decay of the fraction~$f(t)$ of orbits following the self-similar
island-around-island structure.
}
\label{fig:zas}
\end{center}
\end{figure}

Generic Hamiltonian systems are neither integrable nor chaotic~\cite{MM74}, but rather 
exhibit a mixed phase space, where regular and chaotic regions coexist.
Each island of regular motion is surrounded by infinitely many chains of smaller islands.
As the same holds for any of these smaller islands a very complex hierarchical phase-space
structure is found for generic Hamiltonian systems,
which is well understood\cite{lichtenberg} and nowadays appears in textbooks on classical
mechanics\cite{saletan}.
The dynamical properties, however, are still unclarified.
The most fundamental statistical quantity for characterizing dynamics is the decay
of correlations in time.
It determines transport properties and is directly related to the distribution
of Poincar\'e recurrences $P(t)$, which is the probability to return to a given region
in phase space with a recurrence time larger than $t$.
This probability decays on average like a power-law\cite{CS81}
\begin{equation}\label{powerlaw}
P(t)\sim t^{-\gamma}\quad,
\end{equation}
due to the trapping of chaotic trajectories in the hierarchically structured
vicinity of islands of regular motion.
The power-law decay is a universal property of Hamiltonian systems.
It has dramatic consequences for transport\cite{transport} (anomalous diffusion) 
and quantum mechanics\cite{quantum} (conductance fluctuations and eigenfunctions),
which sensitively depend on the value of $\gamma$.
The exponent~$\gamma$, as determined from finite time numerical experiments,
seems to be non-universal, varies with system and parameter, and
typically ranges between $1$ and $2.5$\cite{CS81,transport,quantum}.
It is a fundamental question of Hamiltonian chaos, how the exponent~$\gamma$
of the dynamics is related to the structure of the hierarchical phase space.
Recently, it was argued by Chirikov and Shepelyansky that for asymptotically large times
the exponent is independent of the specific system and parameter and is given by
the universal exponent $\gamma=3$\cite{CS99}.
Their arguments are based on the universal presence of critical tori in phase space 
and are supported by a numerical investigation of the kicked rotor at kicking strength $K=K_c=0.97163540631$.
At this parameter value the golden torus is critical, i.e., it can be
destroyed by an arbitrarily small perturbation.
The self-similar vicinity of the critical golden torus [see Fig.~\ref{fig:density}(a)] 
has been studied using renormalization methods\cite{kay83} and the asymptotic value $\gamma=3$ 
for the power-law decay of $P(t)$ was predicted long time ago~\cite{HCM85,alleCS}.
The fact that it has never been observed led to the speculation that the universal decay
should appear for larger times\cite{murray}.
In Ref.~\cite{CS99} a numerical approach allowed to estimate the onset of this decay
in agreement with the presented data for $P(t)$.
In addition to the sticking of trajectories in the vicinity of critical tori, the
trapping of trajectories in island-around-island structures has been studied
\cite{meiss86,zaslavsky}.
Zaslavsky et al.~\cite{zaslavsky} showed that for the kicked rotor at $K=K^*=6.908745$ the phase space
possesses an island-around-island structure of sequence $3-8-8-8\ldots$ [see Fig.~\ref{fig:density}(c)].
They used this self-similarity to derive the trapping exponent $\gamma=2.25$
by renormalization arguments, which was recently supported by the numerics\cite{ZE00}.
In fact, these renormalization approaches for single self-similar phase-space structures 
can be considered as special cases of the more general binary tree model by Meiss and Ott\cite{MO85}.
In this model a chaotic trajectory can at any stage of the tree either go to
a boundary circle (level scaling) or to the island-around-island structure (class scaling).
The universal coexistence of the two routes of renormalization at any stage led to the
exponent $\gamma=1.96$\cite{MO85}.
In contrast, the recent findings claim that just {\it one} of these scalings is
relevant for the trapping mechanism:
While in Ref.~\cite{CS99} it is argued that universally (and in particular for
$K=K_c$) the level scaling should dominate, in Ref.~\cite{zaslavsky} it is claimed that
for $K=K^*$ the class scaling describes the trapping mechanism.
In order to clarify these contradictions, we numerically investigate $P(t)$
for the kicked rotor at $K_c$ and $K^*$ for times larger than in previous studies.
We find strong deviations from the predictions of the renormalization theories
that only consider a single self-similar phase-space structure.
In addition, our numerical approach allows to analyze where chaotic trajectories are trapped in phase space.
For large times the majority of trajectories is {\it not trapped} in those phase-space regions,
that are described by the simple renormalization theories.
We thereby reveal the mistaken assumption of these theories.
In particular for $K=K_c$, the self-similar vicinity of the critical golden torus does not
dominate the trapping mechanism for large times
and thus even in this ideal situation the proposed universal exponent $\gamma=3$ is not found.
For $K=K^*$, the majority of long trapped trajectories does not follow the self-similar
island-around-island structure, which leads to a smaller exponent than predicted.
Although the phase space in both cases is dominated by exactly self-similar structures,
our analysis shows that they do {\it not} dominate the dynamics.
We use the standard map (kicked rotor) defined by
\begin{equation}\label{standardmap}
q_{n+1}=q_n+p_n \,{\rm mod}\,2\pi\qquad p_{n+1}=p_n+K\sin q_{n+1} \,\, ,
\end{equation}
which has a $2\pi$-periodic phase space in $p$ direction.
We concentrate on two parameters:
(i) The dynamics for $K=K_c$ is bounded in $p$ direction by the golden torus,
which is critical~\cite{kay83}.
The route towards the critical golden torus is determined by the principal resonances
given by the approximants of the golden mean $\sigma=(\sqrt{5}-1)/2$
and the scaling has been analyzed in detail~\cite{kay83}.
The dynamics along this route was described by a Markov chain leading to 
$\gamma=3.05$~\cite{HCM85} and alternatively via the scaling of the local diffusion rate
leading to $\gamma=3$~\cite{alleCS}.
(ii) The phase space for $K=K^*$ consists of two small accelerator modes embedded in
an otherwise completely chaotic phase space.
Each mode shows an island-around-island structure of sequence $3-8-8-8\ldots$~\cite{zaslavsky}.
This exact scaling relation was used to predict the exponent $\gamma=2.25$.
\begin{figure}
\begin{center}
    \epsfxsize=8.4cm
    \leavevmode
    \epsffile{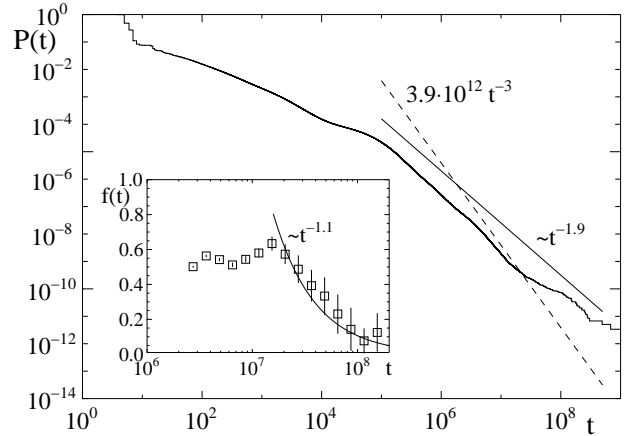}
\caption[fig:tcube]{
Poincar\'e recurrences $P(t)$ (solid) for the standard map with $K=K_c$, 
deviating from the prediction of Ref.~\cite{CS99} (dashed).
The inset shows the decaying fraction $f(t)$ of trajectories following the route of renormalization
towards the critical torus.
}
\label{fig:tcube}
\end{center}
\end{figure}
In order to check the predictions for $P(t)$, in case (i) we start several long trajectories
initially located near the unstable fixed point~$(q,p)=(0,0)$.
We measure the times~$\tau$ for which an orbit stays close to the critical torus by monitoring
successive crossings of the line $p=0$, as was also done in Ref.~\cite{CS99}.
In case (ii) we start many trajectories at four different regions in the chaotic part
of phase space\cite{remark1}.
Whenever a chaotic trajectory is trapped to one of the island structures,
it follows the dynamics of the accelerator mode and jumps to the neighboring unit cell
in $p$ direction.
We measure the time $\tau$ it continuously jumps one unit cell per iteration in the same
direction.
From the set of trapping times~$\tau$ one determines the fraction $\tilde{P}(t)$ of orbits with $\tau\ge t$.
This quantity decays with the same power-law exponent as the Poincar\'e recurrences $P(t)$ and was
chosen for numerical convenience.
The total computer time corresponds to (i) $15\cdot10^{12}$ and (ii) $8\cdot10^{12}$ iterations of
the standard map.
We have checked if our statistical data for large times are sensitive to the unavoidable
finite numerical precision, by comparing data for double ($\approx$16 significant digits)
and quadruple ($\approx$32 digits) precision.
We found no difference and present a combination of both data sets in Fig.~\ref{fig:tcube}.
In Fig.~\ref{fig:tcube} we compare our numerical findings for $P(t)$ for case (i)
with the prediction $P(t>10^7)=3.9\cdot10^{12}\,t^{-3}$ extracted from Ref.~\cite{CS99}.
This power law  is not compatible with our data, even though we are in the time regime,
where it should be observable according to Ref.~\cite{CS99}.
For $10^5\le t\le 10^9$ we rather see an exponent $\gamma\approx1.9$\cite{uppercurve}.
For case (ii) with $K=K^*$, we find a power-law decay of $P(t)$ with $\gamma=1.85$ 
(Fig.~\ref{fig:zas}) for various starting conditions~\cite{remark1} contradicting the 
renormalization prediction $\gamma=2.25$~\cite{zaslavsky}.
Our numerical results raise the question why the renormalization theories predict the wrong exponents. 
This is particularly surprising since the structure of the phase space for the specific parameters
$K_c$ and $K^*$ is dominated by one exactly self-similar hierarchy.
All these approaches rely on the assumption that this single self-similar phase-space structure
also dominates the long-time trapping of chaotic trajectories.
We will now check this assumption.
To this end we calculate the density in phase space for trajectories, that are trapped for long times.
For case~(i) we show two examples in Fig.~\ref{fig:density}(a,b).
Figure~\ref{fig:density}(a) shows a trajectory of length $t\approx5\cdot10^7$ that
follows the route to the critical golden torus up to the principal resonance with winding number $55/144$.
This is consistent with the renormalization theory according to the data presented in Ref.~\cite{CS99}.
In contrast, the trajectory shown in Fig.~\ref{fig:density}(b) approaches the critical torus
only up to the resonance $3/8$ and is predominantly trapped around a non-principal resonance.
This trajectory is not captured by the renormalization theory, since it has the same
length as the trajectory in Fig.~\ref{fig:density}(a) and therefore should be trapped around the
$55/144$ resonance or one of its neighbors.
In order to quantify this observation we introduce
the fraction $f(t)$ of trajectories with trapping time~$t$ that follow the route of renormalization.
Numerically, we determine $f(t)$ by considering trajectories with trapping times in an interval
around~$t$. We classify these trajectories, as was done in Fig.~\ref{fig:density},
according to their phase-space densities.
For $t>2\cdot 10^7$ this fraction decreases to zero (Fig.~\ref{fig:tcube}, inset).
This clearly demonstrates that for large times
the majority of trajectories is {\it not trapped} in the self-similar phase-space regions
approaching the critical golden torus.
The assumption of the renormalization theories leading to $\gamma\approx3$ is violated and thus
they are not applicable for predicting $P(t)$. 
From the ratio of the predicted $P(t)\sim t^{-3}$ for the trajectories trapped in the
self-similar phase-space structure and the observed $P(t)\sim t^{-1.9}$
we expect their fraction to decay as $f(t)\sim t^{-1.1}$ for large times.
This is confirmed by our numerical data in the inset of Fig.~\ref{fig:tcube}.
We find that the majority of trajectories is trapped around non-principal
resonances, which is in agreement with the binary tree model~\cite{MO85}.
This is in strong contrast to the conclusions of Ref.~\cite{CS99}
that are based on the computation of exit times from the vicinity of unstable fixed points
of principal resonances.
As for the investigation of local diffusion rates\cite{ruffo96}
also the analysis of the mean exit time yields information only about trajectories trapped
in the region studied.
One cannot conclude, however, that these trajectories dominate the global trapping mechanism,
as is convincingly shown by our numerics.
We carry out the same analysis for case~(ii), $K=K^*$.
In Figure~\ref{fig:density}(c,d) we show the phase-space densities of two long trajectories.
Although both trajectories have the same length, only
the trajectory in Fig.~\ref{fig:density}(c) follows the self-similar island-around-island structure,
while the trajectory shown in Fig.~\ref{fig:density}(d) is trapped around other islands.
The fraction~$f(t)$ of trajectories following the route of renormalization decays to zero
for large times (Fig.~\ref{fig:zas}, inset).
This decay is well described by the estimate $f(t)\sim t^{-2.25}/t^{-1.85}\sim t^{-0.4}$,
i.e., the ratio of the predicted $P(t)\sim t^{-2.25}$ and the observed $P(t)\sim t^{-1.85}$.
This shows that the renormalization theory for the island-around-island structure is not 
capable of explaining $P(t)$.
It should be noted that this difference is not caused by the effect, that the finite precision
of $K^*$ eventually leads to a breakdown of the self-similarity on very small scales.
In conclusion, our analysis shows that even in the presence of an exactly self-similar
phase-space structure it is not sufficient to describe the trapping mechanism of
chaotic trajectories by only this structure, as was recently claimed
in the literature.
We find that additional island structures may dominate the trapping mechanism for large
times and thus affect the power-law decay of $P(t)$.
Our analysis supports qualitatively the tree model by Meiss and Ott, which allows for
the coexistence of two routes of renormalization at any stage.
Quantitatively, we find slightly smaller exponents.
It remains an open question if there exists a universal asymptotic exponent
for the trapping of chaotic trajectories in Hamiltonian systems.

We thank R.~Fleischmann for helpful discussions.
M.W. acknowledges financial support by an EMBO fellowship.


\begin{thebibliography}{1}
%
\bibitem{MM74} L.~Markus and K.~R.~Meyer, {\em Generic Hamiltonian Dynamical Systems 
are neither Integrable nor Ergodic}, Memoirs of the American Mathematical Society, 
No. 114 (American Mathematical Society, Providence, RI, 1974).
%
\bibitem{lichtenberg} A.~J.~Lichtenberg and M.~A.~Lieberman, {\em Regular and Chaotic Dynamics},
Appl. Math. Sciences~38, 2nd ed., (Springer-Verlag, New York, 1992);
J.~D.~Meiss, Rev.~Mod.~Phys. {\bf 64}, 795 (1992).
%
\bibitem{saletan} J.~V.~Jos\'e and E.~J.~Saletan, {\em Classical Dynamics},
Cambridge University Press, (Cambridge, 1998).
%
\bibitem{CS81} B.~V.~Chirikov and D.~L.~Shepelyansky,
in {\em Proceedings of the IXth Intern. Conf. on Nonlinear
Oscillations, Kiev, 1981} [Naukova Dumka {\bf 2}, 420 (1984)]
(English Translation: Princeton University Report No. PPPL-TRANS-133, 1983);
C.~F.~F.~Karney, Physica {\bf 8} D, 360 (1983);
B.~V.~Chirikov and D.~L.~Shepelyansky, Physica {\bf 13} D, 395 (1984);
P.~Grassberger and H.~Kantz, Phys. Lett. {\bf 113} A, 167 (1985);
Y.~C.~Lai, M.~Ding,  C.~Grebogi, and R.~Bl\"umel, Phys.~Rev. A {\bf 46}, 4661 (1992).
%
\bibitem{transport} T.~Geisel, A.~Zacherl, and G.~Radons, Phys.~Rev.~Lett. {\bf 59}, 2503 (1987);
R.~Fleischmann, T.~Geisel, and R.~Ketzmerick, Phys.~Rev.~Lett. {\bf 68}, 1367 (1992);
M.~F.~Shlesinger, G.~M.~Zaslavsky, and J.~Klafter, Nature {\bf 363}, 31, (1993);
G.~M.~Zaslavsky, {\it L\'evy Flights and Related Phenomena in Physics}, (Springer, Berlin, Verlag, 1995);
G.~Zumofen and J.~Klafter, Phys. Rev. E {\bf 59}, 3756, (1999).
%
\bibitem{quantum} Y.~C.~Lai, R.~Bl\"umel, E.~Ott, and C.~Grebogi, Phys.~Rev.~Lett. {\bf 68}, 3491 (1992);
R.~Ketzmerick, Phys.~Rev.~B {\bf 54}, 10841 (1996);
A.~S.~Sachrajda, R.~Ketzmerick, C.~Gould, Y.~Feng, P.~J.~Kelly, A.~Delage, and Z.~Wasilewski,
 Phys.~Rev.~Lett. {\bf 80}, 1948 (1998);
G.~Casati, I.~Guarneri, and G.~Maspero,  Phys.~Rev.~Lett. {\bf 84}, 63 (2000);
R.~Ketzmerick, L.~Hufnagel, F.~Steinbach, and M.~Weiss, Phys. Rev. Lett. {\bf 85}, 1214 (2000);
B.~Huckestein, R.~Ketzmerick, and C.~Lewenkopf, Phys. Rev. Lett. {\bf 84}, 5504 (2000);
L.~Hufnagel, R.~Ketzmerick, and M.~Weiss, Europhys. Lett., {\bf 22}, 264, (2001);
A.~P.~Micolich {\it et al.}, Phys. Rev. Lett. {\bf 87}, 036802, (2001).
%
\bibitem{CS99} B.~V.~Chirikov and D.~L.~Shepelyansky, Phys.~Rev.~Lett. {\bf 82}, 528 (1999).
%
\bibitem{kay83} R.~S.~MacKay, Physica~D {\bf 7}, 283 (1983).
%
\bibitem{HCM85} J.~D.~Hanson, J.~R.~Cary, and J.~D.~Meiss, J. Stat. Phys. {\bf 39}, 327 (1985).
%
\bibitem{alleCS} 
B.~V.~Chirikov, Lect. Notes Phys. {\bf 179}, 29 (1983);
B.~V.~Chirikov, in {\em Proceedings of the International Conference on Plasma Physics,
Lausanne, Switzerland, 1984} (Commission of the European Communities, Brussels, Belgium, 1984), 
Vol.~2, p.~761;
B.~V.~Chirikov and D.~L.~Shepelyansky, in {\em Renormalization Group}, 
edited by D.~V.~Shirkov, D.~I.~Kazakov, and A.~A.~Vladimirov
(World Scientific, Singapore, 1988), p.~221.
%
\bibitem{murray} N.~W.~Murray, Physica~D {\bf 52}, 220 (1991).
%
\bibitem{meiss86} J.~D.~Meiss, Phys. Rev. A {\bf 34}, 2375 (1986).
%
\bibitem{zaslavsky} G.~M.~Zaslavsky, M. Edelman, and B.~A.~Niyazov, Chaos {\bf 7}, 159 (1997).
%
\bibitem{ZE00} G.~M.~Zaslavsky and M. Edelman, Chaos {\bf 10}, 135 (2000).
%
\bibitem{MO85} J.~D.~Meiss and E.~Ott, Phys. Rev. Lett. {\bf 55}, 2741 (1985);
J.~D.~Meiss and E.~Ott, Physica~D {\bf 20}, 387 (1986).
%
\bibitem{remark1} We have started trajectories randomly placed on a line $p=$const
away from the accelerator modes (upper curve in Fig.~\ref{fig:zas}).
In order to increase statistics for large times, we have started
randomly placed trajectories in 3~different small boxes close to the accelerator mode
in positive direction.
In principle, the asymptotic decay of $P(t)$ might depend on the initial box.
We find, however, that this is not the case, as all four curves are identical for times
$t>2\cdot10^3$.
%
\bibitem{uppercurve} We also studied Poincar\'e recurrences for trajectories approaching
the critical torus from the other side in the same way as in Ref.~\cite{CS99}.
In the range from $10^8<t<10^9$ we find a very slow decay, such that $P(t=10^9)$
is more than three orders of magnitude bigger than the prediction $P(t)=3.98\cdot10^{13}t^{-3}$
from Ref.~\cite{CS99}.
%
\bibitem{ruffo96} S.~Ruffo and D.~L.~Shepelyansky, Phys.~Rev.~Lett. {\bf 76}, 3300 (1996).
%
\end{thebibliography}
\end{document}